\documentstyle[12pt,aasms4]{article}
\def\etal{{\it et al.\ }}

\def\putplot#1#2#3#4#5#6#7{\begin{centering} \leavevmode
\vbox to#2{\rule{0pt}{#2}}
\includegraphics{#1}

\end{centering}}

\begin{document}

\title{Morphology of Star Formation Regions in Irregular Galaxies}

\author{Noah  Brosch\altaffilmark{1}}
\affil{Space Telescope Science Institute \\ 3700 San Martin Drive \\
Baltimore MD 21218, U.S.A.}

\and

\author{Ana Heller and Elchanan Almoznino }

\affil{The Wise Observatory and 
the School of Physics and Astronomy \\ Tel Aviv University, Tel Aviv 69978,  
Israel}
 
\altaffiltext{1}{On sabbatical leave from the Wise Observatory and 
the School of Physics and Astronomy,
Raymond and Beverly Sackler Faculty of Exact Sciences,
Tel Aviv University, Tel Aviv 69978, Israel}
\date{}


\begin{abstract}
The location of HII regions, which indicates the locus of present star 
formation in galaxies, is analyzed for a large collection of 110
irregular galaxies (Irr) imaged in H$\alpha$ and nearby continuum.
The analysis is primarily by visual inspection, although a two-dimensional
quantitative measure is also employed. The two different analyses yield
essentially identical results.
HII regions appear preferentially at the edges of the light distribution, 
predominantly on one side of the galaxy, contrary to what is expected from
stochastic self-propagating star formation scenarios.  This peculiar
distribution of star forming regions cannot be explained by
a scenario of star formation triggered by an interaction with 
extragalactic gas, or by a strong one-armed spiral pattern.

\end{abstract}


\section{Introduction}
The star formation (SF) is a fundamental process in the evolution of galaxies
and is far from being well understood. The SF is usually characterized by the
initial mass function (IMF) and the total SF rate (SFR),
which depends on many factors such as the density
of the interstellar gas, its morphology, its metallicity, {\it etc.}
Generally, four major factors drive star formation in galaxies: 
large scale gravitational instabilities, cloud compression 
by density waves, compression in a rotating galactic disk due 
to shear forces, and random gas cloud collisions. 
In galaxies with previous stellar
generations additional SF triggers exist, such as shock waves from
stellar winds and supernova explosions.  
In dense environments, such as clusters of galaxies and
compact groups, tidal interactions, collisions with other galaxies,
ISM stripping, and cooling flow accretion probably play some role
in triggering the SF process. The triggering mechanisms were
reviewed recently by Elmegreen (1998).

While ``global'' phenomena play a large 
part in grand design spirals, random collisions of interstellar clouds
have been proposed as one explanation for dwarf galaxies with bursts of SF.
Due to their small size, lack of strong spiral pattern, and sometimes
solid-body rotation ({\it e.g.,} Martimbeau \etal 1994, Blok \&
McGaugh 1997), the star formation
in dwarf galaxies cannot be triggered by compression from gravitational 
density waves or by disk shear.
Therefore, understanding SF in dwarf galaxies should be
simpler than in other types of galaxies, because the number of possible
trigger mechanisms is reduced.

The H$\alpha$ emission from a galaxy measures its ongoing SFR
(Kennicutt 1983, Kennicutt \etal 1994).  
The blue  luminosity of a galaxy measures its 
SF integrated over the last $\sim 10^9$ yrs 
(Gallagher \etal 1984).  The red continuum radiation originates
both from relatively young stars which already evolved into red
giants and super-giants, and from a large population of aged 
low-mass stars, if previous SF episodes took place.

Hunter \etal (1998) tested a set of SF predictors on
two small samples of dwarf galaxies, one observed by them and another
derived from de Blok (1997). They found that the ratio of   
HI surface density to the critical density for the appearance of ring
instabilities did not correlate with the
star formation, but that the stellar surface brightness did. From this,
they concluded that possibly some stellar energy input provides the
feedback mechanism for star formation. Brosch \etal (1998) confirmed
that the strongest correlation among a number of parameters tested
on a sample of Virgo cluster dwarf irregular galaxies was between the 
average H$\alpha$ surface brightness and the mean blue surface brightness. 
This is similar to the findings of Phillipps \& Disney (1985) for
spiral galaxies, where in a sample of 77 spiral galaxies from
Kennicutt \& Kent (1983) a correlation was found between the total 
H$\alpha$ emission (expressed as specific SFR or as H$\alpha$ equivalent 
width) and the average blue surface brightness.

On the level of individual HII regions in dwarf irregular galaxies, 
Heller \etal (1998: HAB98) showed that a correlation exists also between 
the H$\alpha$ line flux and the red continuum flux underneath the region, 
measured with the same aperture as the line flux. We emphasize that
in this case the correlation is between local quantities, not for
overall galactic properties. These correlations indicate that the
regulation of SF in Irr's is local and by the existing stellar population. 
The self-regulated evolution of dwarf galaxies has recently been
modeled by Andersen \& Burkert (1997).

We concentrate here on samples of late-type dwarf irregular galaxies 
(DIGs) in the Virgo 
cluster (VC) and elsewhere in the nearby (within 100 Mpc) Universe.
The reason for selecting DIGs is to limit the number of possible SF
trigger mechanisms; DIGs are devoid of large-scale SF triggers, as 
explained above.   We take advantage of the availability of net H$\alpha$
line images to determine the general pattern of the distribution of HII 
regions over the irregular galaxies. As far as we could ascertain,  
such a study of a large sample of irregular galaxies was never published.
Previous attempts to classify DIGs were {\it e.g.,} by Sandage \& 
Binggeli (1984) for low surface brightness (LSB) objects, by Loose 
\& Thuan (1985) for BCDs, and by Patterson \& Thuan (1996) for DIGs.
All used broad-band images to perform the classification.
 
We do not study here the morphology of the galaxies, as reflected
by their light distribution on broad-band images, but rather the
morphology of their ensemble of HII regions. Our goal is to
gain some insight on the star-forming properties of this class of
galaxies. The question of the spatial distribution of HII regions 
in irregular galaxies has been studied previously by Hodge (1969) 
for seven nearby objects, and by
Hunter \& Gallagher (1986) for a larger sample of galaxies.

\section{The samples}

The primary sample consists of 52 DIGs in the VC 
 with HI measurements from Hoffman {\it et al.} (1987, 
1989). The sample was constructed in order to enable the detection
of weak dependencies of the star formation properties on the
hydrogen content
and on the surface brightness. We selected two sub-samples
by surface brightness; one represents a high surface brightness (HSB) group
and is classified as either BCD or anything+BCD, and another represents
a low surface brightness (LSB) sample and includes only ImIV or ImV 
galaxies. The uniform morphological classification, which
bins the DIGs in the HSB or LSB groups, is exclusively from
Binggeli \etal (1985, VCC). 
The galaxies were observed at the Wise Observatory (WO)  
from 1990 to 1997, with CCD imaging through broad bands
and narrow H$\alpha$ bandpasses in the rest frame of each galaxy. 
The discussion of all observations and 
their interpretation is the subject of other papers (AB98, HAB98). We 
restrict the discussion here to the localization of the star-forming
regions on the broad-band, or continuum light images of the DIGs.
In some Virgo galaxies we did not detect H$\alpha$ emission; these
objects have been omitted in Table 1 leaving 30 DIGs from our
combined Virgo sample.

The two Virgo cluster samples are augmented here by 83 irregular
galaxies for which data were collected from the literature.  Not
all these objects are dwarfs but all appear to be, or are
classified as, irregular galaxies; we call them here DIGs
and are not strict in qualifying an object as ``dwarf''. Three objects
of the extended sample appear in two references; these objects
have been classified independently and have two entries in Table 1.
The total number of classifications is thus 124, but only 110 different
objects have been considered.

We inspected H$\alpha$ and broad-band or red continuum
images from Strobel \etal (1991), Miller \& Hodge (1994), McGaugh
\etal (1995), van Zee (1996),  Marlowe \etal (1997), Martin (1997), 
Hilker \etal (1998),  and Gavazzi \etal (1998). 
The images from van Zee include objects analyzed in her PhD thesis 
(van Zee 1996)
and some galaxies from an unpublished comparison sample. The objects were
selected to be LSB DIGs and were checked
not to show obvious signs of interaction on the Palomar Sky Survey plates.
The objects from van Zee contribute 27 galaxies to the extended
sample.
The objects studied by Gavazzi \etal (1998) and included here consist of eight
galaxies classified as Irr, most in the A1367 cluster.
The Hilker \etal (1997) object is an LMC-like galaxy in the Fornax cluster.
Ho II was studied in detail by Puche \etal (1992); we used the published
images for the present classification.
The 12 galaxies studied by Marlowe \etal (1997) are classified as either 
amorphous or blue compact, are intrinsically faint ({\it i.e.,} dwarf),
and are nearby. The sample of Martin (1997) contains a heterogeneous assemblage 
of star-forming dwarf galaxies. Three objects with spiral morphology (N2537, 
VII Zw403, and N4861) from her list were excluded from the present analysis,
leaving 12 galaxies to be considered here. Additionally, four objects 
were added from the study of dwarf galaxies in the M81 system
(Miller \& Hodge 1994) and four other from the study of DIGs by Strobel 
\etal (1991). Finally, we included 13 objects from the morphological
study of LSB disk galaxies of McGaugh \etal (1995) which did not
show strong spiral patterns on the published images.

The only restrictions to the inclusion of a galaxy in the extended sample
were that the object should be classified as an irregular galaxy in the 
original publication and that it would have a net H$\alpha$ and an off-line 
image. This resulted in a very heterogeneos collection of irregular 
galaxies; those from Martin (1997) and Marlowe \etal (1997) are
mainly low-luminosity, nearby objects, while those from Gavazzi \etal 
(1998), being at $\sim$70 Mpc, are of high luminosity and would not
be strictly classifiable as ``dwarfs''.  Most
of the galaxies in  Gavazzi \etal were rejected because they were not 
Irr galaxies and so were all the objects analyzed by Hunter \etal (1998), 
which do not have published
images but only azimuthal averages of line and continuum emission. We 
could not use the large H$\alpha$ and continuum image set in Koopman 
(1997) because it contains only spiral and lenticular
galaxies. The entire selection of 110 galaxies classified here is 
listed in Table 1,
where the objects from our Virgo sample are identified by their number in the 
Virgo cluster catalog (Binggeli \etal 1985).

\section{Analysis and results}

The analysis reported here is based primarily on the visual inspection
 of the net H$\alpha$-line
and off-line images of each galaxy. These are usually presented side-by-side
in the original publications,
at the same scale and with enough ``gray-scale stretch'' to allow easy
perception of the HII regions on the net-line image, and of the general
outline of the galaxy in the off-line image. This facilitates the comparison
and the determination of whether the HII regions are distributed mostly
at the edges or near the center of an object. These two cases have been
noted in Table 1 as E for edge and C for center, and are the primary
morphological index used for this classification.

There are a few cases of mixed morphology, which have been so noted in
Table 1. One such example is UGC 7178, from the
primary sample of van Zee (1996), which has four HII regions, two near the 
center and two at the edge and is classified here as C+E. 
Other galaxies do not show preferential SF at
either the center or the edge, but display a $\sim$linear distribution of 
HII regions showing up as a ``spine'' on the galaxy image. These objects
are noted as L=linear, and/or Sp=spine types and are probably related
to the ``cometary'' galaxies noted by Loose \& Thuan (1985). A few 
objects have a
number of HII regions arranged on the (partial) circumference of an
ellipse; these are marked El=ellipse in Table 1. Finally, some Irr's
show a scattering of HII regions and are accordingly
marked D=diffuse. UGC11820 in van Zee's primary sample (van Zee
1996) has its HII regions arranged on $\sim$a spiral
arm. It is possible that this is a case of mistaken classification and the
object is probably a spiral, as listed in  UGC and  
in NED. A similar case may be object 127037 in Gavazzi \etal (1998).

Our classification, by the distribution of the visible HII regions, should 
be compared with that of Patterson \& Thuan (1996), where six classes
of DIGs were distinguished on the basis of broad-band B and I
images. The classifications are an extension of the Loose \& Thuan
(1986) scheme and bin the DIGs into dwarf spirals (dS), nucleated
dwarf irregulars (dInE), dwarf ellipticals with a central ridge (dIrE),
dwarfs with ``asymmetric star formation'' (dIa), objects with randomly
scattered star formation similar to the GR8 galaxy (GR8), and dwarfs
which show a bar with whispy extensions (dIB). The two UGC objects in
common, U300 and U2162, have both been classified by Patterson \&  
Thuan as dIa, while we classify them as E and A, confirming the asymmetry
mention and adding the qualifier that the HII regions tend to be at the 
edges of the  galaxies.

We used a second morphological index to flag a symmetric or 
asymmetric distribution (S/A) of HII regions. A galaxy is labelled asymmetric
(A) in the distribution of its HII regions if these are located predominantly 
on one side of the galaxy. In other words, the label A is assigned if it 
is possible to draw on the continuum or broad-band image of a galaxy a
diameter which bisects it so that most of the HII regions are on
one side of this diameter. If no such diameter seems to exist, the galaxy
is classified as symmetrical (S) in the distribution of its HII regions.
This asymmetry criterion is similar to that used by Hodge (1969), with the
exception that Hodge used the ``reference frame'' of the HII region 
distribution while we used that of the red continuum light.
Note that the secondary classification of the asymmetry is
independent of the primary classification of edge/center/spine described
above. This secondary classifying index is also listed in Table 1. In
some cases the images were too poor to allow a classification. These objects
have a question mark in the table.
The references for the images are listed in Table 2.

We show in Figure 1 examples of the primary and secondary 
classifications using objects from AB98 and HAB98. The top row shows an 
object with edge distribution of HII regions, which is asymmetric. The 
middle row
shows a galaxy where the HII regions are arranged in a linear, spine-like
configuration. The bottom row shows an object with a centrally located,
symmetric distribution of HII regions.

Obviously, a consistent classification requires similar types of display,
contrast of images, {\it etc.} This is not always the case, as some
sources did not provide two images (line and continuum) for a
galaxy, but only one with overlaid contours
for the missing information ({\it e.g.,} Martin 1997 or Marlowe \etal 1997)
and we also did not have control over the display mode. Nevertheless,
the few galaxies appearing in two references allow some measure of confidence in
the classification: N1800, N5253, and II Zw40 are common to the samples
from Marlowe \etal (1997) and Martin (1997); their classification, performed
completely independently on different images, is virtually identical for 
all three cases.

To put the classification on a more ``objective'' and numerical basis,
and to aviod possible biases caused by the tendency to detect
structures when visually inspecting random distributions of dots,
we formed two indices which quantify the degree of concentration of
HII regions (``concentration index''=CI) and the amount of 
asymmetry in the
HII region distribution (``asymmetry index''=AI). These indices are
calculated from counts of HII identified in various regions of the
galaxies' net-line images we inspected and are listed in Table 1. 
The definition
of these two parameters is fairly intuitive and is explained below.

CI is defined as the ratio of the number of
HII regions in the inner part of the galaxy to that in the outer part.
We count the HII regions within half the semi-major axis from the
center, and divide this by one-quarter of the number of HII regions 
external to this region. The one-quarter factor brings the comparison to 
a number per
equal-area basis and is, in fact, a ratio of the number surface density
of HII regions. CI can have values between zero and infinity, as the outer
part of the galaxy may be devoid of HII regions. A value of unity 
represents a uniform distribution of HII regions, while galaxies with
no emission near the center have CI=0. Galaxies with emission localized
in their centers and no emission detected in their outer parts have 
CI=$\infty$; this is represented in the Table as CI=100.

The asymmetry index AI is defined as the ratio of the number of 
HII regions counted
in the ``HII poor'' half of the galaxy to that in the ``HII rich''
area, where the divider is the bisecting diameter selected visually to
show the largest contrast between the two galaxy halfs. AI ranges
from 0 to 1, with unity representing a symmetric distribution of
HII regions. The smaller the value of AI, the more asymmetric is the
distribution of HII regions.

We explain the two-dimensional morphological classification with the 
example of VCC 17,
shown in the top row of images in Fig. 1. VCC 17 has two central HII 
regions and four regions in its outer part. Its CI is therefore 
$\frac{2}{4/4}=2$. A diameter may be drawn on the figure which puts
two HII regions on one side of it and four on the other side. This appears 
to be the most extreme asymmetry, making AI=$\frac{2}{4}$=0.5.

\section{Discussion}

Our choice of irregular galaxies in which to study the patterns of star
formation was quite deliberate. As mentioned in the introduction, these
objects should be devoid of large-scale SF triggering mechanisms such
as density waves or various disk instabilities. Therefore the triggering 
mechanisms should be simpler to disentangle in this kind of objects.
The star formation indicator used here was the H$\alpha$ line emission 
and the distribution of the HII regions 
was checked against the light distribution of either the red continuum,
or any broad-band images supplied by the authors of a specific paper.

The summary statistic for the distribution among types and the
symmetric or asymmetric morphologies is presented in Table 3. The galaxies
with mixed morphologies have been counted once in each bin, thus the total
number of cases listed in the table is larger than the number of actual 
galaxies inspected. Galaxies with incomplete or dubious classification,
which have only a question mark in the S/A column of Table 1, have been 
excluded from the statistic. Table 3 shows that most galaxies have an
E-type distribution of HII regions. Specifically, we find that  
about half of all  classifications are E and A. The only
other bin populated by a significant amount of irregular
galaxies is C and S ($\sim$one quarter of all classifications) and the 
other bins are essentially empty. Most galaxies, which were selected 
only to be irregulars based on their appearance on broad-band images, 
form preferrentially their HII regions in their
outer regions (type A has $\sim$2/3 of the cases). The objects 
in which the distribution 
of HII regions appears symmetrical are mostly those where it is also central.
This is very similar to what Loose \& Thuan (1985) found for BCDs.

We show the distribution of the two morphological indices in Figures 2
and 3 (the latter shown to emphasize the behavior of the morphological 
indices for CI$\simeq$0).
It is clear that there is a strong dychotomy, because of the objects with
essentially central H$\alpha$ emission which make up the rightmost
part of the plot (AI$\simeq\infty$). The distribution of the objects 
with inner {\bf and}
outer HII regions is shown in Figure 4. The figure shows the dominance 
of the highly asymmetric coverage of DIGs by HII regions, with most 
cases concentrating at AI$\leq$0.5. This implies that in most DIGs
one half of the galaxy has twice or more the number of HII regions
than the other half.

Gerola \& Seiden (1978) proposed that the  mechanism regulating the SF 
in Irr's is the stochastic self-propagating
SF (SSPSF). Their simulations, as well as the more recent ones
by Jungwiert \& Palous (1994), produce preferentially flocculent or
grand-design spirals. If such a mechanism operates in DIGs
it should produce an expanding SF wave which would engulf the 
entire galaxy, or at least those regions with suitable ISM density. 
Unfortunately, the 3D simulations of SSPSF relevant to DIGs (Comins
1983, 1984) do not show ``snapshots'' of the SF proceeding with time 
through the galaxy. Such plots could have been analyzed in a similar 
manner to the galaxy images and possibly some constraint could have been
derived. The few papers which do show such plots ({\it e.g.,}
Gerola \etal  1980) have too few figures to make their analysis statistically
significant. Note also that in large, slowly-rotating disks, like
the LMC, the SSPSF tends to produce mostly long filaments of
young stars which show no special preference of galactic location
(Feitzinger \etal 1987).

If the galaxy is small, 
and a number of SNs explode off its center, it is possible in principle
to have a compression wave travel through the gas and form stars in
suitable location while escaping from the galaxy in places where the
ISM is thin or altogether absent.  It is difficult to estimate the 
likelihood of such a
mechanism but it is possible to examine it in a well-resolved 
object. The nearby DIG Ho II was studied intensively 
by Puche \etal (1992). A comparison of their H$\alpha$ and off-line
images shows that Ho II forms stars near its center. The HI 
synthesis map indicates that the H$\alpha$ emission originates
either at the interfaces between large holes in the HI distribution
or in the small HI holes. Thus this case also does not argue in favour 
of the SSPSF forming stars asymmetrically, or at the edges of a galaxy.

If the ISM in an irregular galaxy is preferentially aligned 
with its long dimension,  a spine of HII regions could form by
the SSPSF mechanism. Similarly, if the galaxy is in a symmetrical
gravitational potential well, its ISM could concentrate at the
bottom of the potential and begin forming stars there in a C
configuration.

Is there a mechanism of SF which would produce predominantly lop-sided
regions of SF at the edge of a galaxy ? The question was posed recently 
for the LMC
by de Boer \etal (1998), who proposed   a mechanism to explain the
 observed distribution  of giant SF structures lining  the
edge of the LMC. They postulate that the interaction between the LMC
gas and gas in the halo of the Milky Way causes gas compression followed 
by star formation. The SF takes place at the interface between the two 
gas distributions, where the interaction beweeen the LMC gas and the MW
gas occurs. The rotation of the LMC moves the regions
with newly formed stars away from the place of formation, the star formation
ceases, and the newly formed stars simply age. 

We tested this possibility with the objects in our Virgo samples which have 
measurements of H$\alpha$ and underlying red continuum emission for 
individual HII regions. The test is reported in detail in HAB98.
If the mechanism of de Boer \etal (1998) is at work, 
we expect a decrease of the H$\alpha$ line intensity simultaneously with an 
enhancement of the red stellar continuum under the HII region as it ages.
The test we performed involved a comparison of the ratios of H$\alpha$ intensities 
and of the red continua for the brightest and the faintest HII regions
identified in the same galaxy. These ratios were compared for 13 objects
with multiple HII regions and a trend in the opposite direction from 
that expected was found; the more intense the underlying
red continuum, the stronger the line emission is. This argues against
the de Boer \etal (1998) proposition and in favour of a mechanism
regulating the star formation through the existing local stellar population.

McGaugh \etal (1995) mention that the m=1 density wave mode may be possible
in LSB galaxies, giving rise to one-armed spirals. In principle, it
is possible that the excess of irregulars with asymmetric, edge-concentrated
star formation is due to this phenomenon, but we deem it unlikely. The 
reason is that if the
SF process would be driven by a density wave, the one-armed spiral
pattern should be visible not only in the distribution of HII 
regions but also in the red continuum image.

A possibly related phenomenon, of displaced light centers with respect to 
the outermost isophotes in a sample of extremely late-type spirals
was recently reported by Matthews \& Gallagher (1997). The phenomenon
was explained as a consequence of the disk, which is the luminous
galaxy, orbiting in an off-center position within an extended dark
halo (Levine \& Sparke 1998). The question of dark matter (DM) halos
in the context of irregular galaxies has also been discussed by Hunter 
\etal (1998), where the conclusion was that the DM may affect the star 
formation by enhancing the gravitational potential. It is possible that 
such a model could explain also the asymmetric patterns of star formation 
which we reported above, but its exploration, with special emphasis
on local density enhancements by the DM, is beyond the scope of this paper.

\section{Conclusions}

We analysed the distribution of regions where star formation takes
place at present in a sample of 110 irregular galaxies. Our results can be
summarized as follows:

\begin{enumerate}
\item Star formation takes place predominantly at the edges of
dwarf irregular galaxies, mostly to one side of a galaxy.

\item Existing models of star formation in such objects
do not predict such a distribution of star forming regions over
a galaxy. 

\item The proposals of de Boer \etal (1998), of an interaction between 
the galaxy and some surrounding medium which compresses the ISM and
thus enhances the star formation, and of McGaugh \etal (1995),
of strong single-arm spiral patterns in dwarf galaxies which could 
give rise to asymmetric distribution of star-forming regions, can
 probably be rejected. 

\item No good explanation was identified for the peculiar location 
of the HII regions in irregular galaxies.

\end{enumerate}

\section*{Acknowledgements}
NB is grateful for continued support of the Austrian Friends of 
Tel Aviv University and for the hospitality of the Space Telescope
Science Institute, where most of this paper was written. EA is 
supported by a special grant from the 
Ministry of Science to develop TAUVEX, a UV imaging experiment. 
AH acknowledges support from the US-Israel Binational Science 
Foundation and travel grants from the Sackler Institute for Astronomy. 
Astronomical research at Tel Aviv University
is partly supported by a grant from the Israel Science Foundation.
Discussions on this subject with Mario Livio appreciated. We
are grateful for constructive remarks on this subject from Lyle
Hoffman, Crystal Martin, and an anonymous referee. Liese Van Zee 
and Giuseppe Gavazzi
kindly supplied electronic copies of images from their samples of 
galaxies, used for some of the comparisons presented here.

\newpage

\section*{References}
\begin{description}

\item Almoznino, E. \& Brosch, N. 1998, MNRAS, in press (AB98).

\item Andersen, R.-P. \& Burkert, A. 1997, ApJ, preprint.

\item Brosch, N., Heller, A. \& Almoznino, E.  1998, ApJ, 504 (september 10), in press.

\item Binggeli, B., Sandage, A. \& Tamman, G.A. 1985, AJ, 90, 1681.

\item Comins, N.F. 1983, ApJ, 266, 543.

\item Comins, N.F. 1984, ApJ, 284, 90.


\item de Blok, E. \& McGaugh, S.S. 1997, MNRAS 290, 533.

\item de Boer, K.S., Braun, J.M., Vallenari, A. \& Mebold, U. 1998,
A\&A, 329, L49.

\item Gerola, H. \& Seiden, P.E. 1978, ApJ, 223, 129.

\item Gerola, H., Seiden, P.E. \& Schulman, L.S. 1980, ApJ, 242, 517.

\item Gallagher, J.S., Hunter, D.A. \& Tutukov, A.V. 1984, ApJ, 284, 544.

\item Gavazzi, G., Catinella, B., Carraso, L. Boselli, A. \& Contursi,
A. 1998, AJ, 115, 1745.


\item Elmegreen, B.G. 1998, in {\it Origins of Galaxies, Stars, Planets
and Life} (C.E. Woodward, H.A. Thronson, \& M. Shull, eds.), ASP
series, in press.

\item Feitzinger, J.V., Haynes, R.F., Klein, U., Wielebinski, R. \&
Perschke, M. 1987, Vistas in Astr., 30, 243.

\item Heller, A., Almoznino, E. \& Brosch, N. 1998, MNRAS, submitted (HAB98).

\item Hilker, M., Bomans, D.J., Infante, L. \& Kissler-Patig, M. 1997, A\&A,
327, 562.

\item Hodge, P. 1969, ApJ, 156, 847.

\item Hoffman, G.L., Helou, G., Salpeter, E.E., Glosson, J. \& Sandage, A. 
1987, ApJS,  63, 247.

\item Hoffman, G.L., Lewis, B.M., Helou, G., Salpeter, E.E. \& Williams, H.L. 1989, 
ApJS,  69, 65.




\item Hunter, D.A. \& Gallagher, J.S. 1986, PASP, 98, 5.

\item Hunter, D.A., Elmegreen, B.G. \& Baker, A.L. 1998, ApJ, 493, 595.

\item Jungwiert, B. \& Palous, J. 1994, A\&A, 287, 55.


\item Kennicutt, R.C. 1998, ApJ, 498, 541.

\item Kennicutt, R.C. 1983, ApJ, 272, 54.

\item Kennicutt, R.C., Tamblyn, P. \& Congdon, C.W. 1994, ApJ, 435, 22.

\item Levine, S.E. \& Sparke, L.S. 1998, ApJL, in press (astro-ph/9803146).

\item Loose, H.-H. \& Thuan, T.X. 1985 in {\it Star-forming Dwarf
Galaxies} (D. Kunth, T.X. Thuan \& J. Tran Than Van, eds.), Gif sur Yvette: 
Editions Frontieres, p. 73.

\item Marlowe, A.T., Meurer, G.R. \& Heckman, T.M. 1997, ApJS, 112, 285.

\item Martin, C. 1997, ApJ, 491, 561.

\item Martimbeau, N., Carignian, C. \& Ray, J.-R.  1994, AJ, 107, 543.

\item Matthews, L.D. \& Gallagher, J.S. 1997, AJ, 114, 1899.

\item McGaugh, S.S., Schombert, J.M. \& Bothun, G.D. 1995,
AJ, 109, 2019.

\item Miller, B.W. \& Hodge, P. 1994, ApJ, 427, 656.

\item Patterson, R.J. \& Thuan, T.X. 1996, ApJS, 107, 103.

\item Phillipps, S. \& Disney, M. 1985, MNRAS, 217, 435.

\item Puche, D., Westphal, D., Brinks, E. \& Roy, J.-R. 1992, AJ, 103, 1841.

\item Schmidt, M. 1959, ApJ, 129, 243.



\item Strobel, N.V., Hodge, P. \& Kennicutt, R.C. 1991, ApJ, 383, 148.

\item Tresse, L. \& Maddox, S.J. 1998, ApJ, 495, 691.

\item van Zee, L. 1996, PhD thesis, Cornell University.
 
\end{description}
 
\newpage

\section*{Figure captions}
\begin{itemize}
\item Figure 1: Examples of galaxies of different types and morphologies. Each row
shows the galaxy as imaged through the H$\alpha$ filter at left, through the
continuum filter at the center, and in the net-H$\alpha$ line at right. The galaxies
are VCC 17 (E \& A), VCC 1374 (L, Sp \& S), and VCC 10 (C \& S). A 10 arcsec
bar at the upper right corner of each row sets the scale.

\item Figure 2: Distribution of classification indices for DIGs. Each independent
classification is represented by a circle. The points in regions of high
concentration have been ``jiggled'' slightly in an attempt to separate 
points which are very close together. Concentration index values of 
100 represent galaxies 
with CI=$\infty$ ({\it i.e.,} all HII regions in central locations).
 
\item Figure 3: Distribution of classification indices for DIGs. This is an 
expansion of Figure 2 near CI$\simeq$0. The points have {\bf not} been jiggled
in this presentation.

\item Figure 4: Distribution of the asymmetry index among galaxies which have 
CI$\neq\infty$.
Most of the objects concentrate in the low AI bins, indicating a preference 
for asymmetric distribution of HII regions.

\end{itemize}

\newpage
 
 \begin{deluxetable}{cccccc|cccccc}
\tablecaption{Location of HII in irregular galaxies}
\small
\tablehead{\colhead{Galaxy} & \colhead{Type} & \colhead{CI} &
\colhead{S/A.} & \colhead{AI} & \colhead{Ref.}
& \colhead{Galaxy} & \colhead{Type} & \colhead{CI} &
\colhead{S/A.} & \colhead{AI} & \colhead{Ref.}}

\startdata
U300	& E & 0 & A & 0.5&	1& U191		&E& 0 & A & 0.58&2 \nl
U521	& C & 100 & S & 1&	1& U634		&E& 0& A &0	&2 \nl
U2684	& E+C& 100 & A & 1 &	1& U891		&L, Sp& 100 & S&1&2 \nl
U2984	& E, El & 0 & A &0.38&	1& U1175	&E&0& A		&0&2 \nl
U3174	& E & 1.33 & A & 0.4&	1& U2162	&E& 0&  A&0.5 &2 \nl
U3672	& E & 0 & A	&0 &	1&  U3050	&  C+E&1.33& A&0.5&2 \nl
U4660	& D & 0 & A	&0.33&	1& U4762	&E& 0&A	&0.5&2 \nl
U5716	& E, El& 0&A& 0&	1& U5764	& C+E& 0&A& 0&2 \nl
U7178	& C+E& 0.67& A&0.17&	1& U5829	&E& 0&A	&0&2 \nl
UGCA357	& E & 0& A&0&		1& U7300	&E& 0&A	&0.33& 2 \nl
HARO43	& L, Sp & 2& S&1&	1& U8024	&E& 0& S?&0.29&2 \nl
U9762	&D? & -& ?& -	&	1& U9128	&E& 0& S&0	&2 \nl
U10281	&E& 0&A	&0&		1& DDO210	&   C& 100&S&0	&2 \nl
U11820	&E & 0.87& A&0.58&	1& N1427A	&E& 0.2& A&0.11&4 \nl
97073	&E, El& 0.31& A& 0&	3& V0017	&E& 2&A	&0.5&7 \nl
97079	&E& 0& A&0	&	3& V0169	&E& 0&A	&0.5& 7 \nl
97087	&L, Sp& 1&A&1	&	3& V0217	&E&0& A	&0.5&7 \nl
97138	&E&0.66& A&0.25	&	3&V0328		&E& 0&A & 0.5	&7 \nl
108085	&E& 0&A	&0	&	3&V0350		&  C?& 0& S& 0&7 \nl
127037	& E, El&0.57& A&0.20&	3&V0477		&E& 0&A	&0&7 \nl
160086	&    C& 0.57&S	&0.75&	3& V0530	&    C& 100&A&0&7 \nl
160139	&E& 0&A	&0.22	&	3&V0826		&E&0& A	&0.5&7 \nl
HARO 14 &   C & 100&A&1	&	5& V0963	&    C& 100&S&0&7 \nl
N625 	&    C & 100&S	&1&	5& V1455	& C & 100 &A&0	&7 \nl
N1510 	&    C& 100&S&1 &	5&V1465		&C+E& 2.5 & A &0&7 \nl
N1705	&  C& 100&S&1	&	5&V1468		&E& 0&A	&1&7 \nl
N1800 	&   C+E& 1& S&0	&	5& V1585	&E& 0&A	&0&7 \nl
N2101 	&  C+E& 0.33&S	&0.23&	5& V1753	&E& 0&A	&0&7 \nl
II Zw40 &  C & 1 & S	& 1 &	5&V1952		&E& 0&A&0	&7 \nl
N2915 	&  C+E& 100&A	&0.66&	5&V1992		&E& 0&A	&0&7 \nl
N3125 	&  C+E& 1&A	&1&	5&V2034		&D?& - & ? & - & 7 \nl
N3955 	&  C& 100&A?&1	&	5& V0010	&C&100&S &1&	8 \nl
N4670	&    C& 100&S	&1&	5& V0144	&C&100&S&1	&8 \nl
N5253	&    C& 100&S	&1&	5&V0172		&C+E&0&A&0	&8 \nl
N1569	&    C& 1&S&0.33&	6&V0324		&C&100&S&1	&8 \nl
N1800	&E& 100&A & 1&		6& V0410	&C&100&S&1	&8 \nl
II ZW 40&   C& 1&S	&0.25 &	6& V0459	&E&0&A	&1&8 \nl
N2363	&    C& 100&S	&1&	6& V0513	&C&100&S&1	&8 \nl
I ZW 18	&  C& 4&S	&0&	6&V0562		&C& 100 & S & 1 &8 \nl
M82	&  C& 100&S	&1&	6& V0985	&E&100&A&1	&8 \nl
N3077	&    C& 100&S	&1&	6& V1179	&C&100&S&1	&8 \nl
SEX A	&E& ?&A	&?&		6&V1374		&L, Sp &2&A&1	&8 \nl
N3738	&   C& 100&A	&1&	6&V1725		&L, Sp &4&A&0	&8 \nl
N4214	&L, Sp& 0.5&S?&	0.33& 	6&V1791		&E&0&A	&0.5&8 \nl
N4449	&L, Sp&?& A&?&		6&Ho II		&C&1	&S&1	&9 \nl
N5253	&  C &100& S &1&	6&DDO 47	& D&1.28& S & 0.89&10 \nl
Ho I	& E& 0.13&A& 0.35&	11&  Sex B	& E &0.36& A & 0.5&10	\nl
Ho IX	& E &0&A &0.20&		11 &DDO 167 & C	& 25&A & 0.66& 10	 \nl
IC 2547 & E & 0&A& $\sim$0.5&11 &DDO 168 & L, Sp & 1& S & 0.71&10\nl
M81dB	&L, Sp & 0.25&S	& 0.33& 11 & DDO 187 & E	&0& A & 0.5&10\nl
F415-3  & C & 2 & A & 0 & 12  	& F469-2 & C+E & 2 & A & 0 & 12 \nl
F561-1 & E & 1.14 & A & 0.75 & 12 & F563-V1 & D & 100 & S & 1 & 12 \nl
F611-1 & E & 0 & A & 0 & 12 	& U12695 & E & 0.5 & A & 0.14 & 12 \nl
F562-V2 & C+E & 0 & A & 0.5 & 12 & F746-1 & E & 0.36 & A & 0.87 & 12 \nl
U5709 & C & 100 & S & 1 & 12 	& F568-1 & C+E & 0.28 & S & 1 & 12 \nl
F583-5 & L, Sp & 0 & S & 0.75 & 12 & F585-3 & C & 0.8 & S & 0.66 & 12 \nl
U5675 & C & 0 & S & 1 & 12 \nl
\enddata

\tablecomments{The second column gives the typology for the location of 
the HII regions with
the following coding: E=egde, C=center, S=scattered, D=diffuse,
El=on (part of) ellipse, L=Linear, Sp=spine. The numerical concentration index
CI, defined in the text, is given in column 3. The symmetry classification  is
given in column 4, while the asymmetry classifier AI is given in column 5.}
\end{deluxetable}

\newpage

 \begin{deluxetable}{rlc}
\tablecaption{References for classification images}
\small
\tablehead{\colhead{No.} & \colhead{Reference}  &
\colhead{Galaxies}   }

\startdata
1 & van Zee 1996 primary 	& 14\nl
2 & van Zee 1996 secondary 	& 13 \nl
3 & Gavazzi \etal 1998 		& 8\nl
4 & Hilker \etal 1998 		& 1 \nl 
5 & Marlowe \etal 1997 		& 12 \nl
6 & Martin 1997 		& 12 \nl
7 & Heller \etal 1998 		& 17 \nl
8 & Almoznino \& Brosch 1998 	& 13 \nl
9 & Puche \etal 1992 		& 1 \nl
10 & Strobel \etal 1991 	& 4 \nl
11 & Miller \& Hodge 1994 	& 4 \nl
12 & McGaugh \etal 1995 	& 13 \nl \hline
 & Total number of images  	& 113 \nl
\enddata


\end{deluxetable}

\newpage
 
 \begin{deluxetable}{cccc}
\tablecaption{Statistics of types and morphology}
\small
\tablehead{\colhead{Distribution} & \colhead{S morphology}  &
\colhead{A morphology} & \colhead{Total} }

\startdata
 E & 5 & 57 & 62 \nl
D & 2 & 1 & 3 \nl
L & 6 & 4 & 10 \nl
C & 32 & 17 & 49 \nl \hline
Total & 45 & 79 & 124 \nl
\enddata

\tablecomments{Galaxies with mixed morphology are counted once in each
bin (double counting).
}

\end{deluxetable}

\newpage

\begin{figure}[htb]
\putplot{/berlin/data1/noah/Papers/DGs/Ana/V17.ps}{1in}{0}{75}{75}{-200}{-240}
\putplot{ /berlin/data1/noah/Papers/DGs/Ana/V1374.ps}{1in}{0}{75}{75}{-200}{-265}
\putplot{/berlin/data1/noah/Papers/DGs/Ana/V10.ps}{1in}{0}{75}{75}{-200}{-290}
\end{figure}

 \newpage

\begin{figure}[tbh]
\vspace{10cm}
\includegraphics{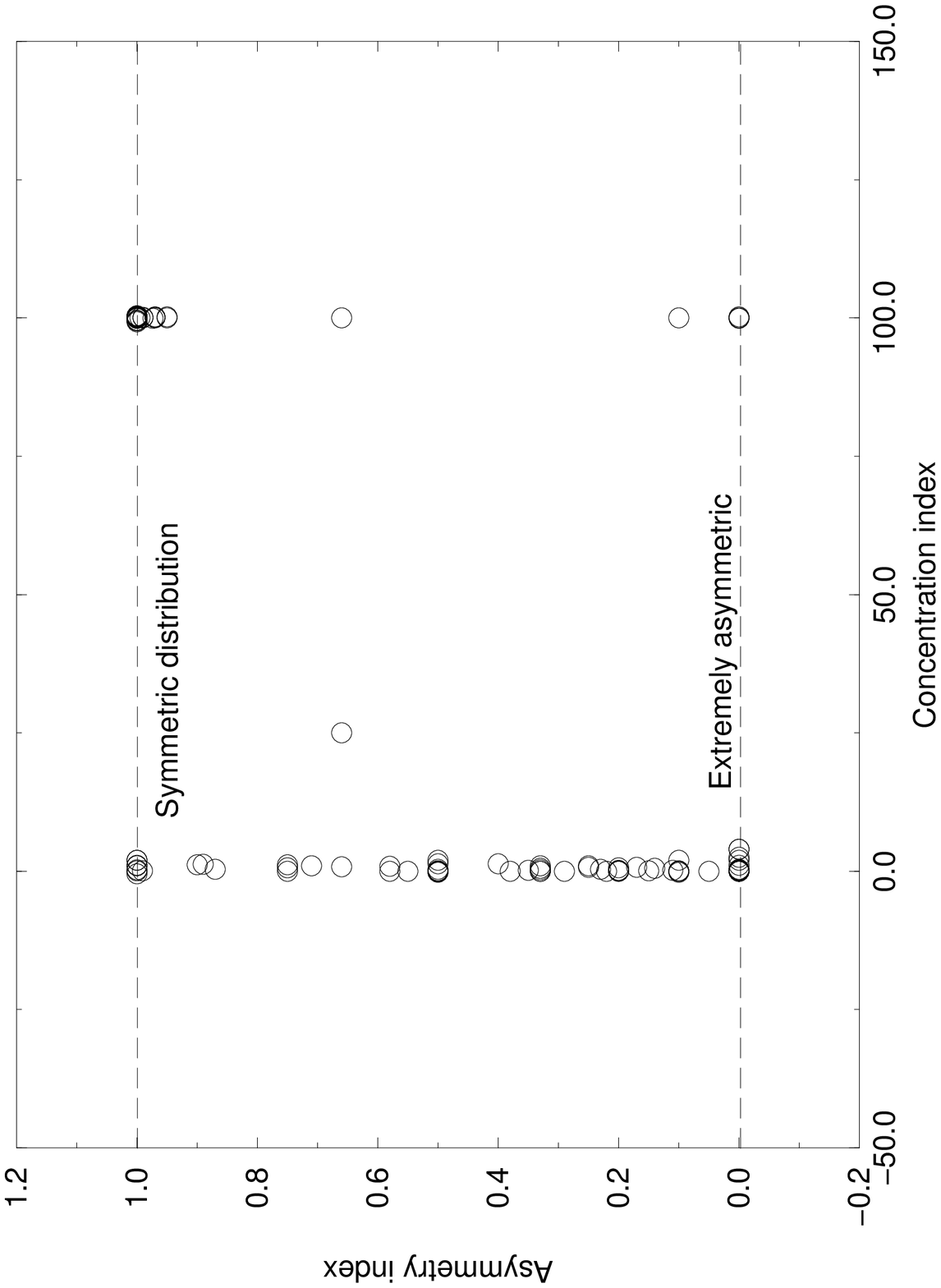}
\end{figure}

 \newpage

\begin{figure}[tbh]
\vspace{10cm}
\includegraphics{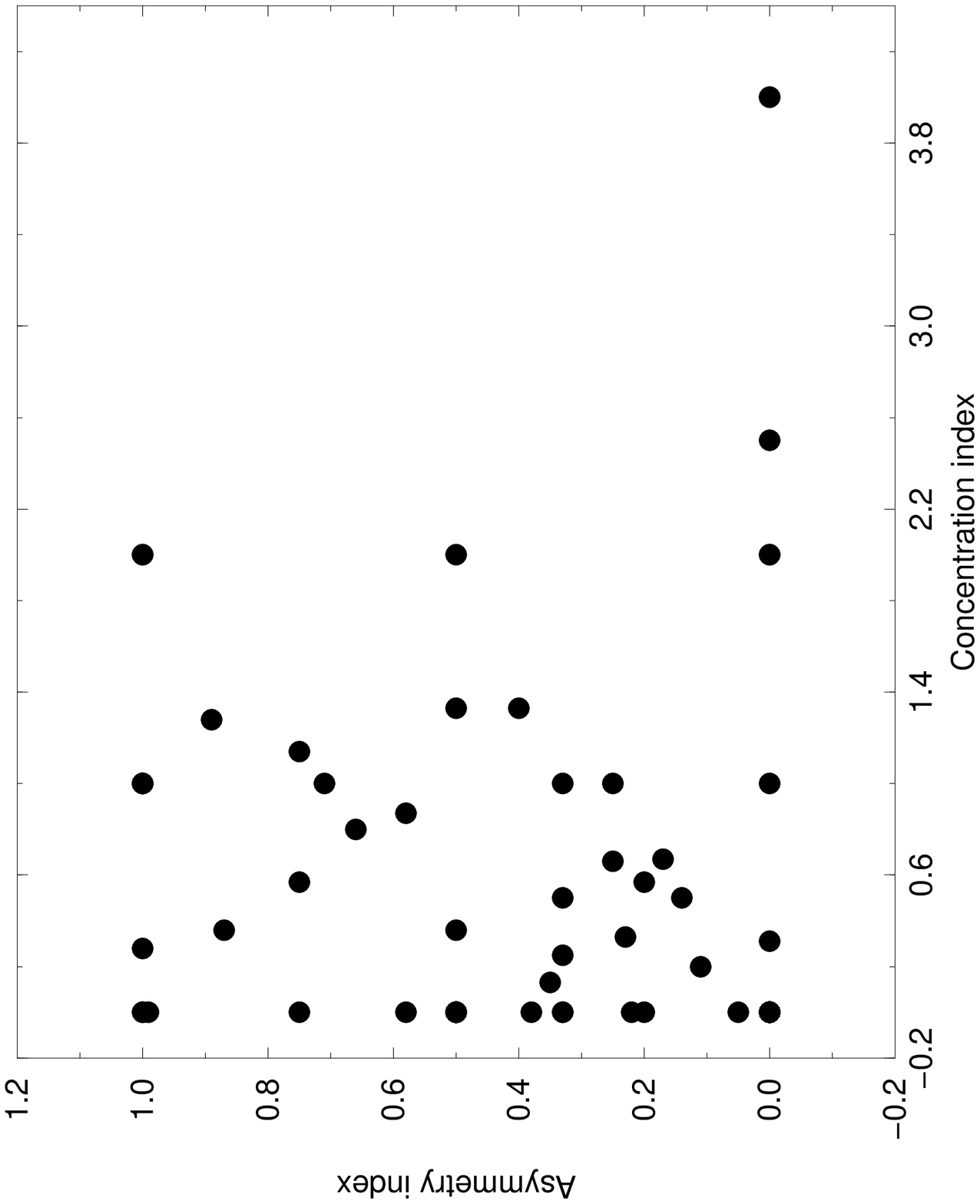}
\end{figure}

 \newpage

\begin{figure}[tbh]
\vspace{10cm}
\includegraphics{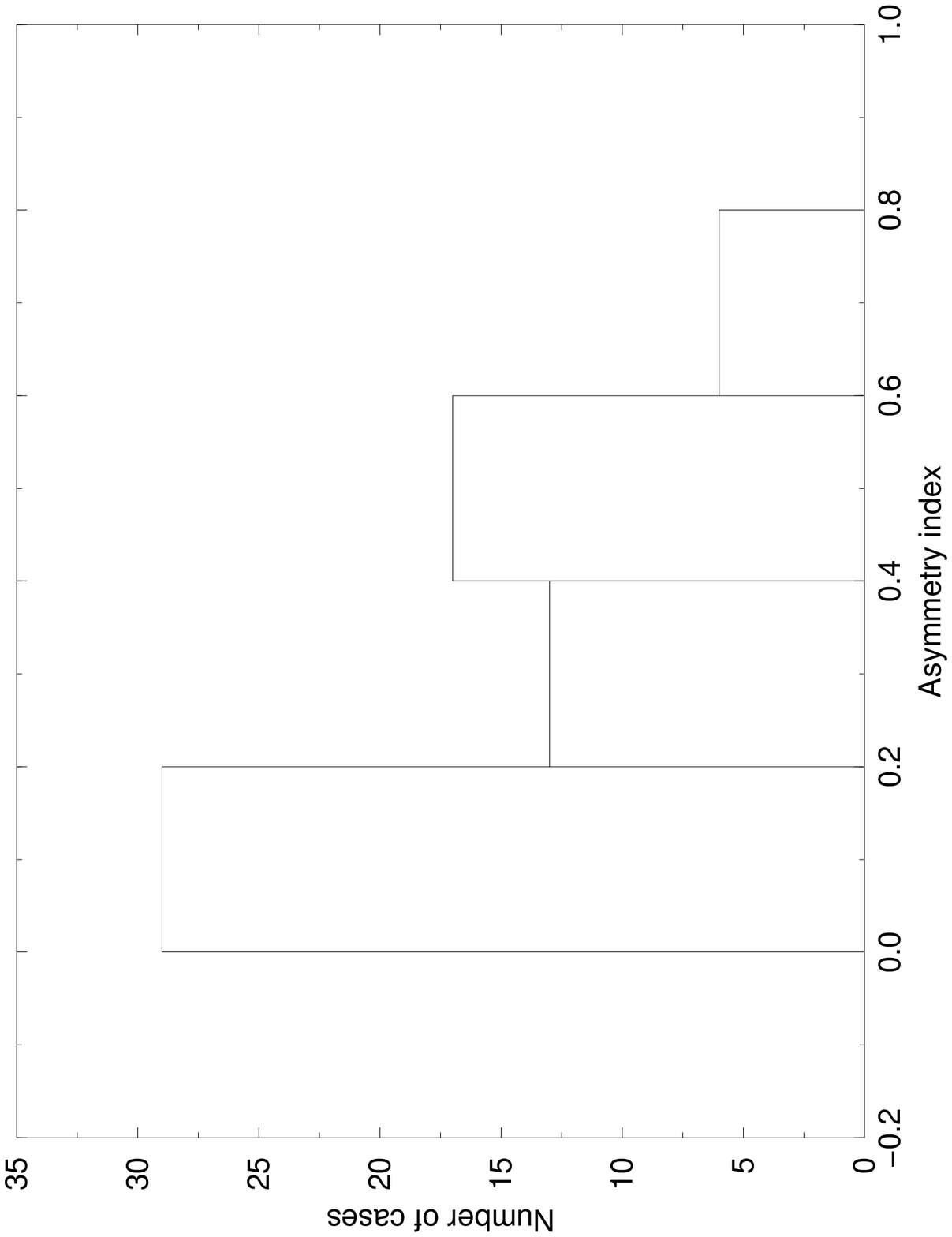}
\end{figure}

\end{document}